\documentclass[10pt]{article} 
\usepackage[top=3cm,bottom=3cm,left=2.5cm,right=3cm]{geometry}

\usepackage{cite}

\usepackage{epsfig}
\usepackage{graphicx}

\usepackage{hyperref}
\hypersetup{colorlinks,citecolor=black,filecolor=black,linkcolor=black,urlcolor=black, 
bookmarksnumbered}

\usepackage{graphicx}
\usepackage{dcolumn}
\usepackage{bm}
\usepackage{amsmath,amsfonts,amssymb,amsthm,amscd,bbm,mathrsfs}
\usepackage{listings}
\usepackage{colortbl}
\usepackage{booktabs}

\newcommand{\gbf}[1] {\mbox{\boldmath${#1}$\unboldmath}}
\providecommand{\norm}[1]{\lVert #1\rVert }

\theoremstyle{definition}

\usepackage{bbm}
\numberwithin{equation}{section}

\title{Synthesis of Spherical 4R Mechanism for  Path Generation using Differential Evolution}
\author{
$^1$F. Pe\~nu\~nuri\footnote{francisco.pa@uady.mx} ,~ 
$^1$R. Pe\'on-Escalante\footnote{rpeon@uady.mx} ,~$^1$C. Villanueva\footnote{cesar.villanueva@uady.mx} ,~
$^2$Carlos A. Cruz-Villar\footnote{cacruz@cinvestav.mx}\\
{\footnotesize \textit{$^1$Facultad de Ingenier\'ia, Universidad Aut\'onoma de Yucat\'an, A.P. 150, Cordemex, M\'erida, Yucat\'an, M\'exico.}}\\
{\footnotesize \textit{$^2$Cinvestav-IPN, Departamento de Ingenier\'ia El\'ectrica, Av. IPN 2508, A. P. 14-740, 07300, M\'exico D.F., M\'exico.}}
}

\date{\today}

\begin{document}
\maketitle
\begin{abstract}
The problem of path generation for the spherical 4R mechanism is solved using the Differential Evolution algorithm (DE).
Formulas for the spherical geodesics are employed in order to obtain
 the parametric equation for the generated trajectory.
  Direct optimization of the objective function gives the solution to the path generation task without prescribed timing. Therefore, there is no need to separate this task into two stages to make the optimization.
   Moreover, the order defect problem can be solved without difficulty by means of manipulations of the individuals in the DE algorithm. Two examples of optimum synthesis showing the simplicity and effectiveness of this approach are included.
\end{abstract}


\section{Introduction}
Spherical mechanisms have a large number of applications (see for example, \cite{Suh1978, Chiang1988, Chablat2003, Lum2006,Hess2007, LeeRussel2009, Kocabas2009, Menon2009, Lehman2010, McDonald2010}), being the spherical 4R mechanism one of the most studied.
Nevertheless, the vast majority of these studies deal with function generation, and there are relatively few studies on the  prescribed path generation task for optimal synthesis of spherical mechanisms \cite{chusun2010}.
The optimal synthesis of spherical mechanisms for the path generation task without prescribed timing, has been addressed in  \cite{Angeles1992}. 
The synthesis procedure presented there consists of two layers of optimization, which are handled separately. In an inner layer, the objective function is minimized over the input angles with fixed linkage dimensions for a given iteration of the outer layer.
The output of this procedure is the set of input angles defining the linkage configurations under which the coupler point attains the closest position from a given point.
On the other hand, the outer layer minimizes the linkage error over the design vector.
In this process, iterations do not stop until some conditions are met.
Recently, two studies have been published regarding the prescribed timing case  \cite{chusun2010, Mullineux20111811}.
In these studies, the Fourier coefficients are used to describe the coupler curve, then the path generation task is done by using the atlas method.

In this study we address the problem by optimizing the structural error which is defined as the sum of the square of the distances between the desired and the generated points.  The generated points are obtained by finding the spherical geodesic between two points.  The optimization is done by using the evolutionary algorithm known as Differential Evolution (DE), that allows us to modify the main operators and thus provide a simple way to deal with the order defect problem.

 


When dealing with the task of path generation without prescribed timing, the order defect problem has to be solved.
The methodology to obtain the parametric equation for the generated trajectory that is proposed here,
 allows the designer to deal with this problem without difficulties.
The optimization process can be done by minimizing a single objective function without the need of making a sequential optimization.

The paper shows in detail how to construct the objective function ($f_{ob}$) for optimal synthesis of spherical mechanisms in both cases: the prescribed timing and the not prescribed timing path generation. We also work out two examples of optimum synthesis by using the proposed methodology.

\section{Mathematical preliminaries}\label{sec2}
In this section, relevant concepts and useful formulas involved in the synthesis of spherical mechanisms are presented.
In particular, a rotation matrix is used to construct the parametric equation for spherical geodesics 
which makes it possible to construct the parametric equation of the generated trajectory.

\subsection{The rotation matrix}
In the study of synthesis of a spherical four bar mechanism one kind of transformation is very important: the one that keeps the  the length of a vector invariant.
It is well known that rotations of the coordinate system, or rotations of a vector, keep the vector length invariant.
Both transformations are carried out by orthogonal matrices ($\mathbf{A}^{\text{T}} \mathbf{A}=\mathbbm{1}$).
If  the coordinate system is rotated, the rotation is known as passive; if the rotation is performed on the vector, the term active rotation is used \cite{Nikravesh1998}.
In the passive case,  the vector remains stationary and rotations are made on the coordinate axes. 
In the active case, the vector is rotated while the coordinate system remains fixed.
Relationship between them is given in Eq. (\ref{rotaM}).

In order to find the rotation matrix,  the generators of the $SO(3)$ group (the set of $3\times 3$ orthogonal matrices with determinant equal to 1) are required.

The generators $ \mathbf{T}_k$ are obtained from the formula:
\begin{equation}
\mathbf{T}_k=-i\left(\frac{\partial {\mathbf{R}}_k}{\partial \theta_k}\right)_{\theta_k=0}\,, \quad k=x,y,z,
\end{equation}
 where ${\mathbf{R}}_k$ represents the passive rotation of an angle $\theta_k$ around the coordinate axis $k$ and $i = \sqrt{-1}$.
Such matrices are:
\begin{eqnarray}\nonumber
{\mathbf{R}}_ x=\left[
\begin{array}{ccc}
 1 & 0 & 0 \\
 0 & \cos \theta_x  & \sin \theta_x  \\
 0 & -\sin \theta_x  & \cos \theta_x 
\end{array}
\right],\\
{\mathbf{R}}_ y=\left[
\begin{array}{ccc}
 \cos \theta_y & 0 & -\sin \theta_y \\
 0 &  1 & 0  \\
 \sin \theta_y &  0 & \cos \theta _y
\end{array}
\right],\\
{\mathbf{R}}_ z=\left[
\begin{array}{ccc}\nonumber
 \cos \theta _z &  \sin \theta _z & 0 \\
  -\sin \theta_z &  \cos \theta_z & 0  \\
 0 &  0 & 1
\end{array}
\right].
\end{eqnarray}
Thus,
\begin{eqnarray}\nonumber
\mathbf{T}_ x=-i\left[
\begin{array}{ccc}
 0 & 0 & 0 \\
 0 & 0  & 1  \\
 0 & -1  & 0 
\end{array}
\right],\\
\mathbf{T}_ y=-i\left[
\begin{array}{ccc}
0 & 0 & -1 \\
 0 &  0 & 0  \\
 1 &  0 & 0
\end{array}
\right],\\
\mathbf{T}_ z=-i\left[
\begin{array}{ccc}\nonumber
 0 &  1 & 0 \\
  -1 &  0 & 0  \\
 0 &  0 & 0
\end{array}
\right].
\end{eqnarray}

Let us consider a passive rotation of an angle $\theta$ around a unit vector
 $\hat{\bf{n}}=\left[n_x\; n_y\; n_z\right]^{\text{T}}$ .
The rotation matrix is then written in terms of the generators as:
\begin{equation}
{\mathbf R}_{\text{passive}}(\theta,\hat{\bf{n}})=\text{exp}\left(i\,\theta \sum_{k=x}^{z}\mathbf{T}_k n_k\right).
\end{equation}
On the other hand the active rotation would be:
\begin{equation}\label{rotaM}
{\mathbf{R}}(\theta,\hat{\bf{n}})={\mathbf R}_{\text{passive}}(-\theta,\hat{\bf{n}}).
\end{equation}
 It is worthwhile to mention that there are other forms to describe the rotation of a vector, an interesting discussion can be found in \cite{Angeles1991}.


\subsection{Spherical geodesics}

%
%
%

It is assumed that with a point $(x,y,z)$ on the space there is a vector associated to it, namely, the vector from the origin to the  point $(x,y,z)$. As it is customary, we will use the same symbol to represent both, it will be clear from the context if we are talking about the point or the vector.

In order to construct the geodesics on the spherical surface, let us consider two points ${\mathbf h}_1$ and ${\mathbf h}_2$ on the surface of a sphere of unit radius.
The parametric equation for the spherical geodesic ${\mathbf r}_{g12}$ from ${\mathbf h}_1$ to ${\mathbf h}_2$ is constructed by rotating the vector ${\mathbf h}_1$ an angle $\cos^{-1}({\mathbf h}_1\cdot{\mathbf h}_2)$ around of the normalized  cross product vector ${\mathbf h}_1 \times {\mathbf h}_2$ denoted by $\hat{\mathbf{n}}_{h12}$.


\begin{equation}\label{geodesicEq}
{\mathbf r}_{g12}(\theta;{\mathbf h}_1,{\mathbf h}_2)={\mathbf R}(\theta,\hat{{\mathbf n}}_{h12}){\mathbf h}_1,
\end{equation}
where $\theta$ is the parameter of the trajectory varying from $\theta=0$ up to $\theta=\cos^{-1}({\mathbf h}_1\cdot{\mathbf h}_2)$.


\section{Kinematic modeling}
\begin{figure}[h]
\begin{center}
\includegraphics[scale=0.35]{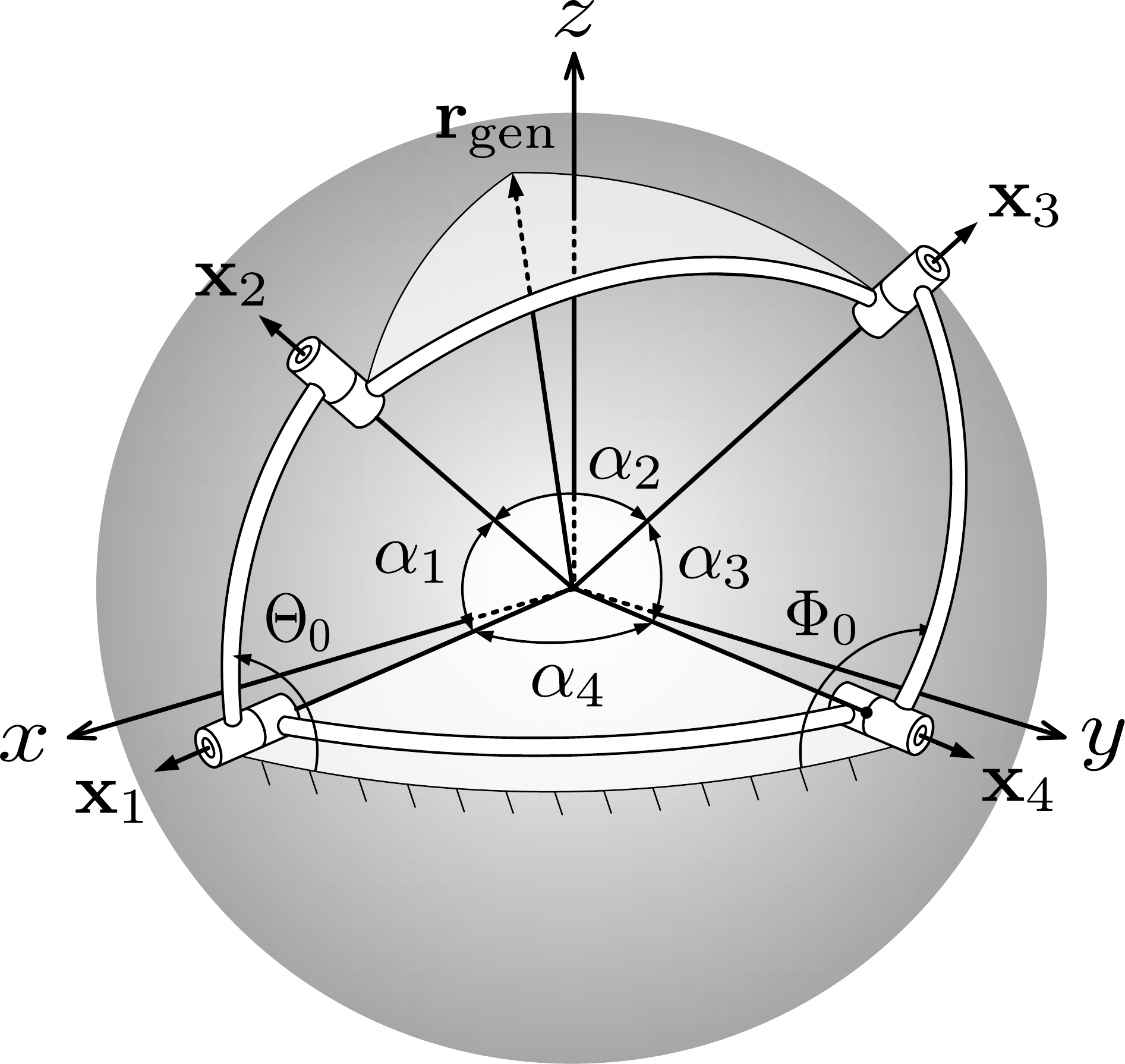} 
\caption{Variables involved in  the Spherical four-bar mechanism.}
\label{figure1}
\end{center}
\end{figure}

Let  ${\mathbf x}_1,~{\mathbf x}_2,~{\mathbf x}_3$ and ${\mathbf x}_4$  be four arbitrary points on the spherical surface. 
We will consider the input link as the geodesic connecting the points ${\mathbf x}_1$ and ${\mathbf x}_2$, the coupler link as the geodesic connecting the points ${\mathbf x}_2$ and ${\mathbf x}_3$, the output link as the geodesic connecting the points ${\mathbf x}_3$ and ${\mathbf x}_4$, finally the fixed link will be the geodesic connecting ${\mathbf x}_4$ and ${\mathbf x}_1$. 
Assuming that we are in a unit sphere, 
the lengths of the links will be the angles between the vectors defining the links. We will call  $\alpha_1$, $\alpha_2$, $\alpha_3$, $\alpha_4$
the lengths of the input link, the coupler link, the output link and the fixed link respectively. Fig. \ref{figure1} shows the aforementioned variables.

Now, let us consider  that the input link is rotated from the assembly position by an angle $\theta$.  The parametric trajectory for the extreme of such a link, ${\bf r}_2(\theta;{\mathbf x}_1,{\mathbf x}_2)$, will be obtained by rotating the vector ${\mathbf x}_2$ around the vector ${\mathbf x}_1$ an angle $\theta$,
\begin{equation}\label{r2}
 {\mathbf {r}} _2(\theta;{\mathbf x}_1,{\mathbf x}_2)={\mathbf R}(\theta,{\mathbf x}_1){\mathbf x}_2.
\end{equation}
 Such a rotation will generate in the output link a rotation given by the angle $\phi(\theta)$. The parametric trajectory for the extreme of the output link is written as
\begin{equation}\nonumber 
 {\bf r}_3(\theta;{\mathbf x}_1,{\mathbf x}_2,{\mathbf x}_3,{\mathbf x}_4) \equiv {\bf r}_3 (\phi(\theta);\vec{{\mathbf x}}), ~\text{with} ~\vec{{\mathbf x}}=({\mathbf x}_1,{\mathbf x}_2,{\mathbf x}_3,{\mathbf x}_4).
\end{equation} 
  It is given by 
\begin{equation}\label{r3}
 {\mathbf r}_3(\phi(\theta);\vec{{\mathbf x}})={\mathbf R}(\phi(\theta),{\mathbf x}_4){\mathbf x}_3.
\end{equation} 
The dependence of equation (\ref{r3}) on ${\mathbf x}_3$ and ${\mathbf x}_4$ is clear, the dependence on ${\mathbf x}_1$ and ${\mathbf x}_2$ comes from $\phi(\theta)$ (in fact, to be precise we should write $\phi(\theta;\vec{{\mathbf x}})$, although it is not written to avoid unnecessary notation).

Notice that the parameter $\theta$ is not the angle between the geodesic associated to the fixed link  and the geodesic associated to the input link but the rotation angle of the $\mathbf x_2$ vector around the  $\mathbf x_1$ vector. Similarly $\phi(\theta)$ is not the angle between the geodesics associated to the fixed link  and the output link but the rotation angle of the ${\bf x}_3$ vector around the ${\bf x}_4$ vector.

The angle $\phi(\theta)$ is obtained by requiring  the coupler link to have a constant length. In other words, 
\begin{equation}\label{coplercondition}
{ \mathbf r}_2(\theta;{\mathbf x}_1,{\mathbf x}_2)\cdot {\mathbf r}_3(\phi(\theta);\vec{{\mathbf x}})={\mathbf x}_2\cdot{\mathbf x}_3=\text{constant}.
\end{equation}
Solutions for $\phi(\theta)$ from Eq. (\ref{coplercondition}) can be obtained either numerically or analytically.
Indeed, the analytical solution can be found in \cite{Cervantes2009}, which in our notation is given by:
\begin{align}\label{phiangleB}
\phi=\, &\Phi_0  - 2 \tan^{-1}\left(\frac{A\pm \sqrt{A^2+B^2-C^2}}{C-B}\right)\text{where}\\
A=\,&\sin \alpha _1 \sin \alpha _3 \sin (\theta + \Theta_0) \nonumber\\
B=\,&\cos \alpha _1 \sin \alpha _3 \sin \alpha _4 -\sin \text{$\alpha _1$} \sin \text{$\alpha _3$} \cos
   \text{$\alpha _4$} \cos(\theta +  \Theta_0) \nonumber\\
 C=\,&\sin \text{$\alpha _1$} \cos \text{$\alpha _3 $} \sin \text{$\alpha _4$} \cos(\theta + \Theta_0) +\nonumber\\
   &\cos \text{$\alpha _1$} \cos \text{$\alpha _3$}
   \cos \text{$\alpha _4$}-\cos \text{$\alpha _2 $}\nonumber,
\end{align}
with $\Theta_0$ being the initial angle  between the geodesics associated to the fixed link and the input link.
$\Phi_0$ is the angle  between the geodesics associated to the fixed link and the output link.

 
\begin{figure}[hbtp]
\begin{center}
\includegraphics[scale=0.35]{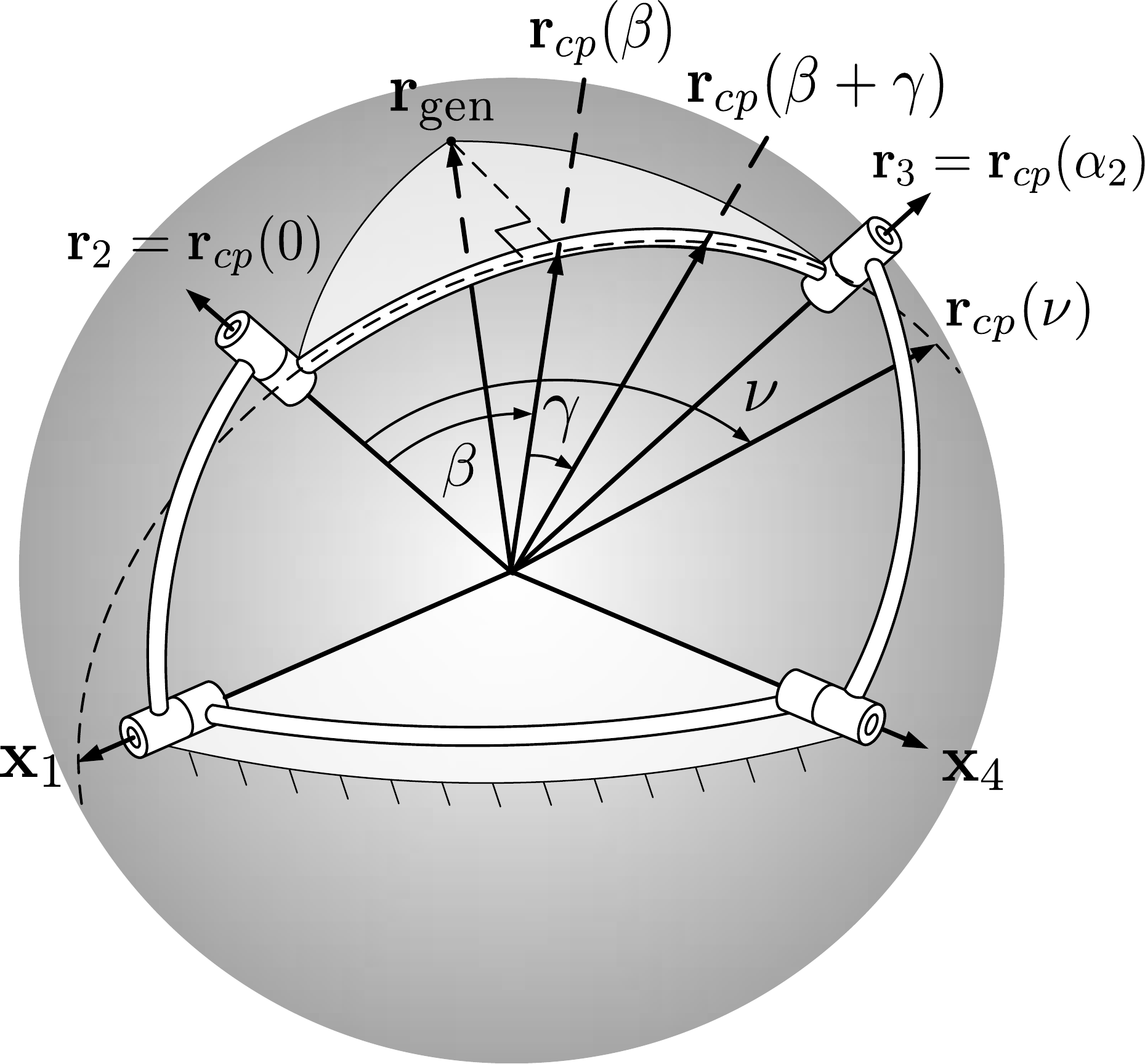} 
\caption{Vectors involved to obtain ${\mathbf r}_{\text{gen}}$. The dependence on $\theta$ and $\vec{{\mathbf x}}$ is not written.}
\label{figure2}
\end{center}
\end{figure} 
 
 In order to obtain the parametric equation for the generated trajectory ${\mathbf r}_{\text{gen}}$ (as it is shown in Fig. \ref{figure2}), it is necessary to find the position vector ${\mathbf r}_{cp}(\theta,\nu;\vec{{\mathbf x}})$ of an arbitrary point on the coupler link (or on the complete arc that contains the coupler link, since $0\leqslant \nu < 2\pi$). This can be done by rotating the vector  ${\mathbf r}_2(\theta;{\mathbf x}_1,{\mathbf x}_2)$ by an angle $\nu$ around the unit vector  $\hat{{\mathbf n}}_{23}$ orthogonal to ${\mathbf r}_2(\theta;{\mathbf x}_1,{\mathbf x}_2)$ and ${\mathbf r}_3(\phi(\theta);\vec{{\mathbf x}})$. 
Eq. (\ref{geodesicEq}) then yields:
\begin{equation}\label{couplerpoint}
{\mathbf r}_{cp}(\theta,\nu;\vec{{\mathbf x}})={\mathbf R}(\nu,\hat{{\mathbf n}}_{23}){\mathbf r}_2(\theta;{\mathbf x}_1,{\mathbf x}_2),
\end{equation}
where 
\begin{equation}
\hat{\bf{n}}_{23}=\dfrac{{\mathbf r}_2(\theta;{\mathbf x}_1,{\mathbf x}_2) \times {\mathbf r}_3 (\phi(\theta);\vec{{\mathbf x}})} {\norm{{\mathbf r}_2(\theta;{\mathbf x}_1,{\mathbf x}_2) \times {\mathbf r}_3(\phi(\theta);\vec{{\mathbf x}})}}. 
\end{equation}
Using Eq. (\ref{couplerpoint}), the  parametric equation for the vector ${\mathbf r}_{\text{gen}}$ is obtained by rotating the vector ${\mathbf r}_{cp}(\theta,\beta + \gamma;\vec{{\mathbf x}})$ an angle $\pi/2$ around the vector ${\mathbf r}_{cp}(\theta,\beta;\vec{{\mathbf x}})$:

\begin{equation}\label{rgenx}
{\mathbf r}_{\text{gen}}(\theta,\beta,\gamma;\vec{{\mathbf x}})={\mathbf R}\left(\pi /2,{\mathbf r}_{cp}(\theta,\beta;\vec{{\mathbf x}}) \right){\mathbf r}_{cp}(\theta,\beta + \gamma;\vec{{\mathbf x}}).
\end{equation}

Since the point ${\mathbf x}_k$ $(k=1,2,3,4)$ is on the unit sphere, two parameters are required to determine
its location. In this work the azimuthal $\varphi$ (latitude) and polar $\eta$ (colatitude)  angles are used to generate the point, hence ${\mathbf x}_k={\mathbf x}_k(\varphi,\eta)$.
Explicitly,
\begin{equation}\label{jointspoints}
{\mathbf x}_k (\varphi _k,\eta _k)= (\cos \varphi _k \sin \eta_k, \sin \varphi _k \sin \eta _k, \cos \eta_k).
\end{equation}
In the case of the prescribed timing situation,  Eq. (\ref{rgenx}) depends on eleven parameters, or on ten parameters if all input angles are given\footnote{In fact, it could be possible to reduce the number of parameters by introducing a frame of reference ($S'$), in which the point for the input link joint attached to the fixed link lies on the $x'$ axis, but then all the desired points need to be changed to the $S'$ frame; this is not a difficult task. However, it is not necessary to introduce other frameworks.}. Those parameters will be the adjustment parameters (or independent variables) for the optimum synthesis problem.


%

\section{Path generation}\label{sec4}

The task of function generation for the spherical mechanism has been widely studied. However there are relatively few%
\footnote{As far as we know, only  \cite{Angeles1992} addresses this problem.} 
articles addressing the task of path generation without prescribed timing.
Here, we present how to construct the objective function for this case.
The optimization of such a function is performed directly (unlike other approaches that consider two nested optimization loops) and the order defect problem can be solved without difficulties when the DE method is used.
\subsection{Prescribed timing}\label{sec41}
In the problem of path generation  $n$ points are given and a mechanism that passes through these points is required. In the case of a continuous path a discretization approach can be used.

When the trajectory is prescribed the input angles for each of the desired points are known, or in the worst scenario, are a function of one --- the $k$-th input angle (typically the first). In such a case 
there will be an extra adjustment parameter, the $k$-th or the first input angle.

The objective function $f_{ob}$ for the structural error is:
\begin{equation}\label{objectivef}
f_{ob}=\sum_{i=1}^n \norm{{\mathbf r}_{di}-{\mathbf r}_{\text{gen}i}}^2,
\end{equation}
where ${\mathbf r}_{di}-{\mathbf r}_{\text{gen}i}$ is the difference between the $i$-th desired point ${\mathbf r}_{di}$ and   the $i$-th generated point ${\mathbf r}_{\text{gen}i}$ which can be obtained using Eq. (\ref{rgenx}).
In principle, the sum of the length of the geodesics connecting  ${\mathbf r}_{di}$ with ${\mathbf r}_{\text{gen}i}$ is what must be minimized, but it is clear that the minimum of Eq. \eqref{objectivef} implies the minimum of the sum of the geodesics connecting the desired and generated points.
This can be seen by calculating the inner product of the vector  ${\mathbf e_i}= {\mathbf r}_{di}-{\mathbf r}_{\text{gen}i}$ with itself, i.e., ${\mathbf e_i}\cdot {\mathbf e_i}=2(1-\cos \delta_i)$. Where $\delta_i$ is the length of the geodesic connecting ${\mathbf r}_{d i}$ with ${\mathbf r}_{\text{gen}i}$.

\subsection{Path generation without prescribed timing}\label{sec42}
Unlike the prescribed timing case, in the nonprescribed case the input angles  are not known. Therefore, it is necessary to introduce as many extra parameters as points to generate.
Formally, this does not represent a problem, as Eq. (\ref{rgenx}) deals with it, but the optimization turns out to be more complicated, because the number of adjustment parameters increases.
Let us analyze the situation for the case of $n$ desired points.

Defining 
\begin{equation*}
f_{ob}(\theta _1,\theta_2,\dots,\theta_n,\beta,\gamma,\varphi_1,\varphi_2,\varphi_3,\varphi_4,\eta_1,\eta_2,\eta_3,\eta_4 ) \equiv f_{ob}(\theta _1,\theta_2,\dots,\theta_n,\beta,\gamma,\gbf{\varphi},\gbf{\eta} ),
\end{equation*}
where $\gbf{\varphi}=(\varphi_1,\varphi_2,\varphi_3,\varphi_4)$ and $~\gbf{\eta}= (\eta_1,\eta_2,\eta_3,\eta_4)$.

The objective function is then defined as
\begin{align}\label{objectivef3}
f_{ob}(\theta _1,\theta_2,\dots,\theta_n,\beta,\gamma,\gbf{\varphi},\gbf{\eta} )=&\sum_{k=1}^n \norm{{\mathbf r}_{dk}-{\mathbf r}_{\text{gen}}(\theta _k,\beta,\gamma,\gbf{\varphi},\gbf{\eta})}^2.
\end{align}

So, the design variable vector is
\begin{equation}
{\mathbf X}_D =  \left\lbrace \theta _1,\theta_2,\dots,\theta_n,\beta,\gamma, \varphi_1,\varphi_2,\varphi_3,\varphi_4, \eta_1,\eta_2,\eta_3,\eta_4 \right\rbrace.
\end{equation}

The domain for each of the adjustment parameters in the design variable vector is given as follows
\begin{equation*}
\theta_k \in [0,2\pi); \beta \in [0,2\pi); \gamma \in [-\pi,\pi); \varphi_k \in [0,2\pi); \eta \in[0,\pi].
\end{equation*}

The next step is to minimize the objective function given by the expression (\ref{objectivef3}).  Care must be taken due to the order defect problem. It would be possible to find a mechanism that passes through the desired points, but not necessarily in the required order. Order is imposed because the input link has an increasing rotation. This problem is solved by requiring the input angles to be in ascending order as:
\begin{equation} \label{ordef}
 \theta _1 < \theta_2 \dots< \theta_n. 
\end{equation} 

Now, if an evolutionary algorithm is used to solve the optimization problem, all the variables of the vector ${\mathbf X}_D$ must be randomly chosen. Then it is likely that the $\theta$ angles do not result in ascending order.
To address this, we could penalize the $f_{ob}$ function (which is a common approach in dealing with constraints \cite{Price_Storn2005}). However, if $n$ is relatively big the  penalization approach would not be appropriate, since for practical purposes all individuals would be penalized (the probability of finding one that would not, is $1/n!$). In other words, the evolutionary algorithm would not have individuals to evolve.

A better approach may be found in \cite{Smaili2007}, which consists of discretizing the search space for the $\theta$ angles, and then looking for the first angle in the first partition and so on. However, there is no guarantee that the minimum will lie in the discretized space, as it might happen that two of the best fitting angles are in the same partition. Since each angle is searched in a different partition the minimum would never be found. Another approach ---which is used here--- is to apply the DE method 
(as in \cite{NewVillage2011} ) to manipulate individuals in such a way that satisfies the constraint \eqref{ordef}.


\section{The Differential Evolution Algorithm}\label{sec5}

Algorithms that are inspired by natural evolution are known as evolutionary algorithms. In such methods there is a population which is susceptible of mutation, crossover and selection. 

The main applications of the evolutionary algorithm have been to optimize functions.
 The optimizing function does not need to be a differentiable function and the state space of possible solutions can be disjoint and can encompass infeasible regions \cite{Fogel2000}. This is an advantage since in the synthesis of the spherical mechanism the discriminant in Eq. (\ref{phiangleB})   can become negative. An individual that make the discriminant negative can not be a mechanism that optimize the objective function, then the evolutionary method will start to look for a better individual in another region of the domain.
 Among the evolutionary algorithms, DE is known for its simplicity and for the excellent results it allows.  Below, the original version of the method is outlined  \cite{Price_Storn2005}.

\begin{enumerate}
\item 
The population is described by:
\begin{eqnarray}\nonumber \label{ClassicDE}
\mathbf{P_{\mathbf{x},g}} &=&(\mathbf{x}_{i,g}), ~i=1,...m;~~g=0,...g_{\text{max}}\\
\mathbf{x}_{i;g}&=&(x^j_{i;g}), ~j=1,...D;
\end{eqnarray}
where $D$, $m$ and $g_{\text{max}}$ represent the dimensionality of $\mathbf{x}$, the number of individuals and the number of generations respectively.
\item
Initialization of population:
\begin{equation*}
	x^j_{i;0}=rand^j(0,1)\cdot (b^j_{\text{U}}-b^j_{\text{L}})+b^j_{\text{L}}.
\end{equation*}
Vectors $\mathbf{b}_{\text{U}}$ and $\mathbf{b}_{\text{L}}$ are the parameter limits and
$rand^j(0,1)$ is a random number in $[0,1)$ generated for each parameter.

\item
Mutation:
\begin{equation}
\mathbf{v}_{i;g}=\mathbf{x}_{r_0;g}+F\cdot (\mathbf{x}_{r_1;g}-\mathbf{x}_{r_2;g}).
\end{equation}
$\mathbf{x}_{r_0;g}$ is called the base vector which is perturbed by the difference of other two vectors.

 $r_0, ~r_1,~r_2~\in \{1,2,...m\},~ r_1\neq r_2\neq r_3\neq i$  .
$F$ is a scale factor greater than zero.
\item
Crossover:\\
A dual recombination of vectors is used to generate the trial vector:
\begin{equation}\label{DECrossover}
\mathbf{u}_{i;g}=u^j_{i;g}=\left \lbrace \begin{array}{l} v^j_{i;g}~~\text{if($rand^j(0,1)\leqslant Cr$ or $j=j_{\text{rand}}$)}  \\
                                														x^j_{i;g}~~\text{otherwise.}
 \end{array}\right. 
 \end{equation}
The crossover probability, $Cr \in [0,1]$, is a user-defined value, $j_{\text{rand}} \in [1,D].$
\item
Selection:\\
The selection is made according to
\begin{equation}\label{finClassicDE}
\mathbf{x}_{i;g+1}=\left \lbrace \begin{array}{l} \mathbf{u}_{i;g}~~
\text{if $f(\mathbf{u}_{i;g})\leqslant f(\mathbf{x}_{i;g})$}  \\
                                														\mathbf{x}_{i;g}~~\text{otherwise.}
 \end{array}\right. 
 \end{equation}
\end{enumerate}

The method just described is known as DE/rand/1/bin. There are
variants of it.  For example, when $F$ is chosen to be a random number
the variant is called \textit{dither}.

\section{Example 1. Prescribed timing path generation}\label{sec6}

The example presented in this section consists in finding the spherical four-bar mechanism that passes by sixty four prescribed points. This problem has been studied originally in \cite {chusun2010},
and later in \cite{Mullineux20111811} when the desired points lie  on a unit sphere. From \cite{Mullineux20111811} 
we take the data to be used,  which we reproduce in Table \ref{table64p} for quick reference. 
In the aforementioned references, the problem is solved using the atlas method.


\begin{table}[h]
\caption{Desired points for the prescribed timing path generation example.}
\centering
\scalebox{.85}{
\begin{tabular}{c c c c }
\hline\hline  \\[-2.ex]  
Point number & Point & Point number & Point\\[0.5ex]
 \hline  \\[-2.ex]
			   1 & (0.85737, -0.18481, 0.48037) & 33 & (0.7887, -0.60370, 0.11578) \\
               2 & (0.82985, -0.20167, 0.52030) & 34 & (0.80152, -0.59270, 0.07900) \\
               3 & (0.80241, -0.21996, 0.55478) & 35 & (0.81378, -0.57959, 0.04311) \\
               4 & (0.77567, -0.23967, 0.58389) & 36 & (0.82552, -0.56433, 0.00841) \\
               5 & (0.75011, -0.26056, 0.60785) & 37 & (0.83678, -0.54700, -0.02478) \\
               6 & (0.72607, -0.28244, 0.62693) & 38 & (0.84759, -0.52763, -0.05611) \\
               7 & (0.70381, -0.30515, 0.64152) & 39 & (0.85807, -0.50641, -0.08530) \\
               8 & (0.68352, -0.32833, 0.65193) & 40 & (0.86819, -0.48344, -0.11200) \\
               9 & (0.66533, -0.35185, 0.65844) & 41 & (0.87804, -0.45889, -0.13596) \\
             10 & (0.64933, -0.37537, 0.66144) & 42 & (0.88763, -0.43304, -0.15689) \\
             11 & (0.63559, -0.39867, 0.66115) & 43 & (0.89704, -0.40611, -0.17448) \\ 
             12 & (0.62415, -0.42159, 0.65781) & 44 & (0.90626, -0.37837, -0.18848) \\
             13 & (0.61504, -0.44389, 0.65167) & 45 & (0.91537, -0.35022, -0.19867) \\ 
             14 & (0.60833, -0.46541, 0.64293) & 46 & (0.92433, -0.32193, -0.20481) \\
             15 & (0.60396, -0.48596, 0.63170) & 47 & (0.93322, -0.29396, -0.20667) \\
             16 & (0.60196, -0.50548, 0.61819) & 48 & (0.94196, -0.26667, -0.20400) \\
             17 & (0.60230, -0.52381, 0.60241) & 49 & (0.95052, -0.24048, -0.19667) \\
             18 & (0.60485, -0.54085, 0.58448) & 50 & (0.95885, -0.21581, -0.18441) \\ 
             19 & (0.60959, -0.55656, 0.56448) & 51 & (0.96685, -0.19315, -0.16707) \\
             20 & (0.61637, -0.57081, 0.54244) & 52 & (0.97430, -0.17289, -0.14452) \\
             21 & (0.62500, -0.58363, 0.51844) & 52 & (0.98096, -0.15541, -0.11656) \\
             22 & (0.63530, -0.59489, 0.49248) & 54 & (0.98652, -0.14104, -0.08315) \\
             23 & (0.64700, -0.60456, 0.46467) & 55 & (0.99052, -0.13007, -0.04430) \\
             24 & (0.65989, -0.61259, 0.43507) & 56 & (0.99244, -0.12263, -0.00019) \\
             25 & (0.67370, -0.61896, 0.40378) & 57 & (0.99174, -0.11870, 0.04874) \\
             26 & (0.68811, -0.62363, 0.37093) & 58 & (0.98774, -0.11822, 0.10185) \\
             27 & (0.70293, -0.62652, 0.33674) & 59 & (0.98000, -0.12085, 0.15807) \\
             28 & (0.71789, -0.62759, 0.30133) & 60 & (0.96819, -0.12626, 0.21604) \\
             29 & (0.73274, -0.62678, 0.26500) & 61 & (0.95226, -0.13415, 0.27426) \\
             30 & (0.74737, -0.62404, 0.22800) & 62 & (0.93252, -0.14411, 0.33115) \\
             31 & (0.76167, -0.61933, 0.19059) & 63 & (0.90956, -0.15600, 0.38515) \\
             32 & (0.77544, -0.61256, 0.15307) & 64 & (0.88422, -0.16959, 0.43519) \\[0.5ex]
  \hline\hline
\end{tabular}
}
\label{table64p}
\end{table}

\begin{table}[h]
\caption{Parameter values of the design variable vector for 64 points with prescribed timing.}
\centering
\scalebox{.85}{
\begin{tabular}{c c c c c c c c c c c}
\hline\hline  \\[-2.ex]  
 $\theta_1$&$\beta$&$\gamma$&$\varphi_1$&$\varphi_2$&$\varphi_3$&$\varphi_4$&$\eta_1$&$\eta_2$&$\eta_3$&$\eta_4$\\[0.5ex]
  \hline  \\[-2.ex]
 0.48867   & 0.23066 & 0.47437 & 0.000009 & 0.38828  & 0.19646 & 0.97780 & 1.57081 & 1.46619  & 0.66128 & 1.34474 \\[0.5ex]
  \hline\hline
\end{tabular}

}
\label{var64p}
\end{table}

\begin{table}[h]
\caption{Lengths of the links for the example of 64 points with prescribed timing.}
\centering
\scalebox{.85}{
\begin{tabular}{c c c c}
\hline\hline  \\[-2.ex]  
 Crank link& Coupler link& Oscillator link& Frame link\\[0.5ex]
  \hline  \\[-2.ex]
 0.40142   & 0.82033 & 0.92503 & 0.99484   \\[0.5ex]
  \hline\hline
\end{tabular}

}
\label{lengthsL}
\end{table}

\begin{figure}[hbtp]
\begin{center}
\includegraphics[scale=0.8]{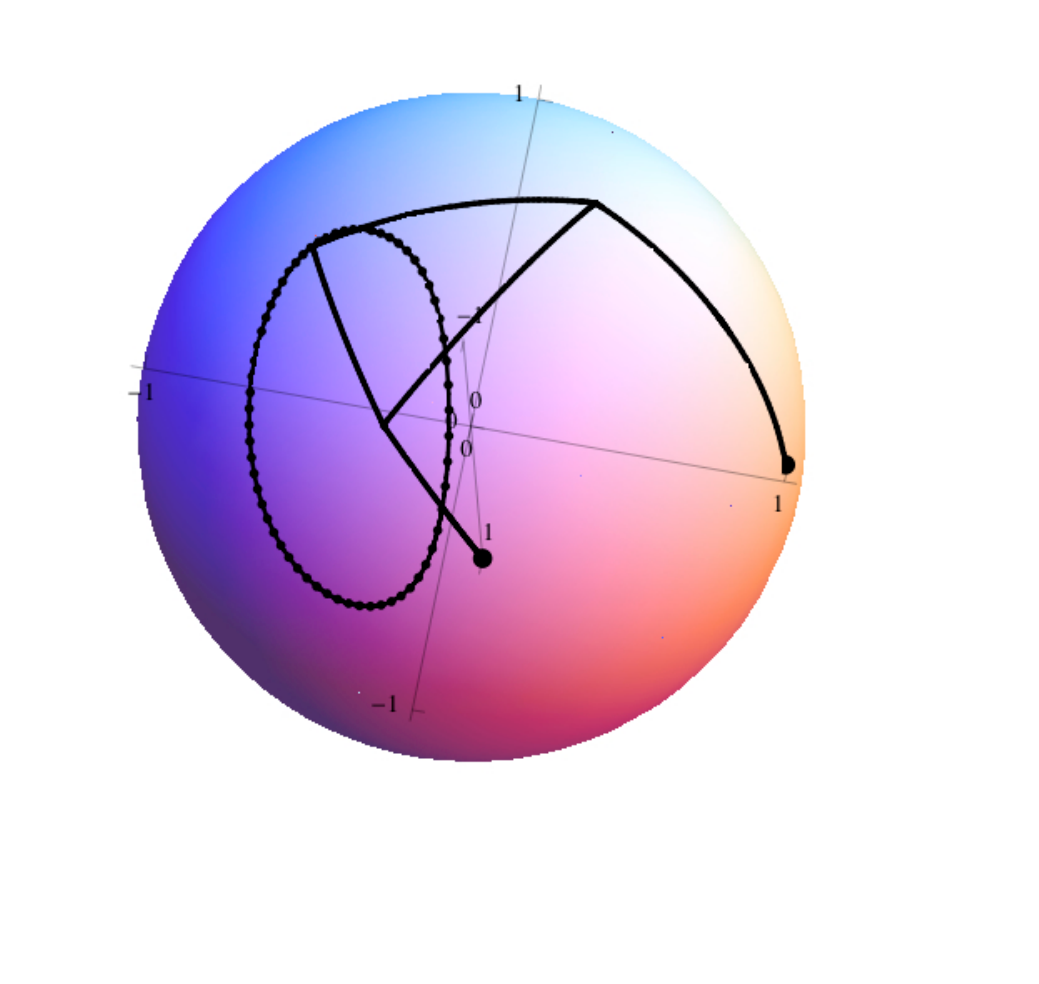} 
\caption{Desired points, generated trajectory and obtained mechanism. Prescribed path generation.}
\label{figure64p}
\end{center}
\end{figure}

In order to optimize the objective function we use the DE/rand/1/bin method with the dither variant.

The first angle $\theta_1$ is not known, therefore it will be an adjustment parameter. The other angles are given by
\begin{equation} \label{preseq}
\theta _ k = \theta _1 + \frac{2\pi}{64}\left(k-1\right); k=2,3,\dots,64.
\end{equation}
The design variable vector is 
\begin{equation}
\mathbf{X}= \left\lbrace \theta _1,\beta,\gamma, \varphi_1,\varphi_2,\varphi_3,\varphi_4, \eta_1,\eta_2,\eta_3,\eta_4 \right\rbrace.
\end{equation}
The objective function is constructed as explained in Section \ref{sec41}. The optimal value for the objective function that was obtained by the DE method is $fob=3.3\times 10 ^{-8}$. In Table \ref{var64p}, we show the optimal design variables.
The DE method was coded in {\sc Fortran}, the  population size was of 100 individuals and a number  of $10\,000$ generations were used. Fig. \ref{figure64p} shows the desired points, the generated trajectory and the obtained mechanism.


\subsection{The Grashof Condition}
Although the Grashof condition \cite{Pierre1998,Cervantes2002,Cervantes2005} can be implemented by a penalty term in the DE method,  we decided to let the method evolve freely. The Grashof condition is verified at the end of the execution program with the purpose of ensuring the mobility of the mechanism in the sense of \cite{Cervantes2002,Cervantes2005}.
Using the data in Table \ref{var64p}, we can calculate the link lengths. For that 
purpose, we must find the points for the joints of the links (using Eq. \eqref{jointspoints})  and then  use the inner product. 
Table \ref{lengthsL} shows the obtained link lengths.  We
can see that the obtained mechanism satisfies the Grashof 
condition.

\section{Example 2. Path generation without prescribed timing}\label{sec7}
\begin{table}[h]
\caption{Angles $\theta$ of the design variable vector for 64 points without prescribed timing.}
\centering
\scalebox{.85}{
\begin{tabular}{c c c c c c c c}
\hline\hline  \\[-2.ex]  
 $\theta_1$&$\theta_2$&$\theta_3$&$\theta_4$&$\theta_5$&$\theta_6$&$\theta_7$&$\theta_8$\\[0.5ex]
  \hline  \\[-2.ex]
2.98039$\times10 ^{-7}$& 0.0977871& 0.195564& 0.293198& 0.390854& 0.488426& 0.58619& 0.684141  \\
\midrule
$\theta_{9}$&$\theta_{10}$&$\theta_{11}$&$\theta_{12}$&$\theta_{13}$&$\theta_{14}$&$\theta_{15}$&$\theta_{16}$\\
 0.782192& 0.880383& 0.978634& 1.07687& 1.17502& 1.27299& 1.37043 &1.4675  \\
\midrule
$\theta_{17}$&$\theta_{18}$&$\theta_{19}$&$\theta_{20}$&$\theta_{21}$&$\theta_{22}$&$\theta_{23}$&$\theta_{24}$\\
1.5642& 1.66089& 1.7576& 1.85419& 1.95116& 2.04815& 2.14514& 2.2422 \\
\midrule
$\theta_{25}$&$\theta_{26}$&$\theta_{27}$&$\theta_{28}$&$\theta_{29}$&$\theta_{30}$&$\theta_{31}$&$\theta_{32}$\\
 2.33934& 2.4366& 2.53394& 2.63134& 2.72876& 2.82621& 2.92382& 3.02148  \\
\midrule
$\theta_{33}$&$\theta_{34}$&$\theta_{35}$&$\theta_{36}$&$\theta_{37}$&$\theta_{38}$&$\theta_{39}$&$\theta_{40}$\\
3.11924& 3.21722& 3.31511& 3.41321& 3.51148& 3.60982& 3.70845& 3.80714  \\
\midrule
$\theta_{41}$&$\theta_{42}$&$\theta_{43}$&$\theta_{44}$&$\theta_{45}$&$\theta_{46}$&$\theta_{47}$&$\theta_{48}$\\
3.90616& 4.0052& 4.10437& 4.20389& 4.30343& 4.40325& 4.50295& 4.60259   \\
\midrule
$\theta_{49}$&$\theta_{50}$&$\theta_{51}$&$\theta_{52}$&$\theta_{53}$&$\theta_{54}$&$\theta_{55}$&$\theta_{56}$\\
 4.70184& 4.80104& 4.89973& 4.99843& 5.09691& 5.19563& 5.29455& 5.39354   \\
\midrule
$\theta_{57}$&$\theta_{58}$&$\theta_{59}$&$\theta_{60}$&$\theta_{61}$&$\theta_{62}$&$\theta_{63}$&$\theta_{64}$\\
 5.49265& 5.59183& 5.69105& 5.7901& 5.88913& 5.98799& 6.08645& 6.18493\\[0.5ex]
  \hline\hline
\end{tabular}

}
\label{var64Angs}
\end{table}

\begin{table}[h]
\caption{Remaining parameter values of the design variable vector for 64 points without prescribed timing.}
\centering
\scalebox{.85}{
\begin{tabular}{c c c c c c c c c c}
\hline\hline  \\[-2.ex] 
 $\beta$&$\gamma$&$\varphi_1$&$\varphi_2$&$\varphi_3$&$\varphi_4$&$\eta_1$&$\eta_2$&$\eta_3$&$\eta_4$\\[0.5ex]
  \hline  \\[-2.ex]
 0.293895& 0.455927& 3.10628$\times 10^{-6}$& 0.301721& 0.358277& 1.01732& 1.60055& 1.31837& 0.456631& 1.35054 \\[0.5ex]
  \hline\hline
\end{tabular}

}
\label{var64Oth}
\end{table}
As before, we address the problem of finding the spherical four-bar mechanism that passes by sixty-four points. But unlike the previous section, we  now deal with the non-prescribed timing case where knowledge of Eq. (\ref{preseq}) is not assumed. The purpose of this example is twofold: first, to obtain the mechanism that accomplishes the desired trajectory; and second, to see whether the method can realize that the angles are equally spaced by the amount of $ 2 \pi / 64$. This is a complicated problem that will test the effectiveness of the proposed methodology.

For this case the design variable vector is given by
 \begin{equation}
\mathbf{X}= \left\lbrace \theta _1,\theta _2,\dots,\theta _{64},\beta,\gamma, \varphi_1,\varphi_2,\varphi_3,\varphi_4, \eta_1,\eta_2,\eta_3,\eta_4 \right\rbrace.
\end{equation}

The construction of the objective function is explained in Section \ref{sec42}. As before, the method used to solve the optimization problem was DE/rand/1/bin, coded in {\sc Fortran} with a population of 300 individuals and a number of 50\,000 generations. The order defect problem was handled as in ref \cite{NewVillage2011}. The optimal value for the objective function is $fob=5.7\times 10^{-6}$. The parameter values for the  $\theta$ angles are shown in Table \ref{var64Angs}, the remaining adjustment parameters are shown in table  \ref{var64Oth}. As in the previous example, the obtained mechanism satisfies the Grashof condition.

In order to see if the obtained angles are in agreement with Eq. (\ref{preseq}) we have calculated the differences 
\begin{equation}
\Delta \gbf{\theta} = \{\theta_{2}-\theta_1,\theta_{3}-\theta_2,\dots,\theta_{64}-\theta_{63}\}.
\end{equation}
This set of differences have a mean value of $\mu = 0.0981734$. Notice that $\mu-2\pi/64=1.3\times 10^{-6}$, so the error by taking $\mu$ instead of $2\pi/64$ is smaller than the accuracy with which the points themselves are given.

We close this section by saying that it is possible to reduce the value of $fob$ by increasing the number of generations. However, this is unnecessary since we can realize that the values of Tables \ref{var64p} and \ref{var64Oth} are (for practical purposes) very similar%
\footnote{Although, we could find another mechanism, since it is well known \cite{Chian1984} that the spherical synthesis has not a unique solution, in our case we find the same solution. 
The fact that the design parameters $\theta_1$, $\varphi_3$, and $~\eta_3$ do not have similar values for both mechanism, is regardless, since what really matters is that the links lengths for both mechanisms are very similar. Also notice that the initial points for the input and output links joints are practically the same for both mechanisms.}
.


\section{Conclusions}\label{sectionconclu}
 By constructing the parametric equations of the spherical geodesics, the generated points of a spherical mechanism can be determined in an easy way and the optimum synthesis can be done in a concise manner. 
 The article shows in detail how the construction of the objective function for the task of path generation without prescribed timing can be done.  Using Differential Evolution  the structural error was optimized for two examples. The first example considered the prescribed timing path generation for sixty four points. The method was capable of finding an optimal value for the objective function of $3.3\times 10^{-8}$.
In the second example, we solved the problem of path generation without prescribed timing for the same sixty four points. The non-prescribed timing case demands the introduction of sixty four additional adjustment parameters for the angles. Hence a total of seventy four parameters are required to construct the objective function. The optimization process was done by  minimizing a single objective function without the need for separate optimization stages as has been done in previous studies. The method was capable to obtain an optimal value of $5.7\times 10^{-6}$ for the objective function and also to obtain the correct prescription with a difference of $10^{-6}$ from the exact value.

\section*{Acknowledgments}

 We thank the financial support from PROMEP, SNI, and CONACYT grant 84060.


\begin{thebibliography}{99}

\markboth{{}}{\textbf{References}}

\bibitem {Suh1978} C. H. Suh, C. W. Radcliffe, Kinematics and Mechanisms Design, John Wiley \& Sons, New York, 1978.

\bibitem{Chiang1988} C. H. Chiang, Kinematics of Spherical Mechanisms, Cambridge University Press, Cambridge, England, 1988.

\bibitem{Chablat2003} D. Chablat, J. Angeles, The computation of all 4R serial spherical wrists with an isotropic architecture, Journal of Mechanical Design. 125 (2003) 275-280.

\bibitem{Lum2006} M.J.H. Lum, J. Rosen, M.N. Sinanan, B. Hannaford, Optimization of a spherical mechanism for a minimally invasive surgical robot: theoretical and experimental
approaches, IEEE Transactions on Biomedical Engineering 53 (2006) 1440-1445.

\bibitem{Hess2007} T. Hess-Coelho, A redundant parallel spherical mechanism for robotic wrist application, Journal of Mechanical Design. 129 (2007) 891-895.


\bibitem {LeeRussel2009} W.-T. Lee, K. Russell, Q. Shen, R. Sodhi, On adjustable spherical four-bar motion generation for expanded prescribed positions, Mechanism and Machine Theory. 44 (2009) 247-254. 

\bibitem{Kocabas2009} H. Kocabas, Gripper design with spherical parallelogram mechanism, Journal of Mechanical Design. 131 (2009) 1-9.

\bibitem{Menon2009} C. Menon, R. Vertechy, M.C. Mark\'ot, V. Parenti-Castelli, Geometrical optimization of parallel mechanisms based on natural frequency evaluation: application
to a spherical mechanism for future space applications, IEEE Transactions on Robotics 25 (2009) 12-23.

\bibitem{Lehman2010} A. Lehman, M. Tiwari, B. Shah, S. Farritor, C. Nelson, D. Oleynikov, Recent advances in the CoBRASurge robotic manipulator and dexterous miniature in vivo robotics for minimally invasive surgery, Proceedings of the Institution of Mechanical Engineers, Part C: Journal of Mechanical Engineering Science. 224 (2010) 1487-1494.

\bibitem{McDonald2010} M. McDonald, S.K. Agrawal, Design of a bio-inspired spherical four-bar mechanism for flapping-wing micro air-vehicle applications, Journal of Mechanisms
and Robotics 2 (021012) (2010) 1-6.

\bibitem{chusun2010} J. Chu, J. Sun, Numerical atlas method for path generation of spherical four-bar mechanism, Mechanism and Machine Theory 45 (2010) 867-879.

\bibitem{Angeles1992}  J. Angeles, Liu. Z, The constrained Least-Square Optimization of Spherical Four-Bar  Path Generators, Journal of Mechanical Design 114 (1992) 394-405.


\bibitem{Mullineux20111811} G. Mullineux, Atlas of spherical four-bar mechanisms, Mechanism and Machine Theory 46 (2011) 1811-1823.

%
%
%
%

%
%

\bibitem{Nikravesh1998} P. Nikravesh, Computer-aided analysis of mechanical systems, Prentice-Hall, New Jersey, 1998.

\bibitem{Angeles1991} J. Angeles, O. Ma, Performance Evaluation of Four-Bar Linkages for Rigid-Body Guidance Based on Generalized Coupler Curves, Journal of Mechanical Design 113 (1991) 17-24.

\bibitem{Cervantes2009} J. Jes\'us Cervantes-S\'anchez, Hugo I. Medell\'in-Castillo,  Jos\'e M. Rico-Mart\'inez, Emilio J. Gonz\'alez-Galv\'an, Some improvements on the exact kinematic synthesis of spherical 4R function generators, Mechanism and Machine Theory 44 (2009) 103-121.

\bibitem{Price_Storn2005} K. V. Price, R. M. Storn, J. A. Lampinen, Differential Evolution: A Practical Approach to Global Optimization, Springer, Berlin, 2005.

\bibitem{Smaili2007} A. Smaili, N. Diab, Optimum synthesis of hybrid-task mechanisms using ant-gradient search method, Mechanism and Machine Theory 42 (2007) 115-130.

\bibitem{NewVillage2011} F. Pe\~nu\~nuri, R. Pe\'on-Escalante, C. Villanueva, D. Pech-Oy, Synthesis of mechanisms for single and hybrid tasks using differential evolution, Mechanism and Machine Theory 46 (2011) 1335-1349.

\bibitem{Fogel2000} David B. Fogel, Evolutionary Computation: Principles and Practice for Signal Processing, SPIE Publications, Bellingham, WA, 2000.

\bibitem{Pierre1998}Andrew P. Murray.,  Pierre M. Larochelle.,   A Clasification Scheme for Planar 4R, Spherical 4R and Spatial RCCC Linkages to facilitate Computer Animation, Proceedings of DETC’98 1998 ASME Design Engineering Technical Conferences September 13-16, 1998, Atlanta, Georgia, USA.

\bibitem{Cervantes2002}  J. Jes\'us Cervantes-S\'anchez , Hugo I. Medell\'in-Castillo,
A robust classification scheme for spherical 4R linkages, Mechanism and Machine Theory 37 (2002) 1145–1163.

\bibitem{Cervantes2005}  Hugo I. Medell\'in-Castillo, J. Jes\'us Cervantes-S\'anchez,
An improved mobility analysis for spherical 4R linkages, Mechanism and Machine Theory 40 (2005) 931–947.


\bibitem{Chian1984}  C. H. Chiang, On the Classification of Spherical Four-Bar Linkages, Mechanism and Machine Theory 19 (1984) 283-287.

\end{thebibliography}
\end{document}